Sokolov V. Y., Korzhenko O. Y.
*Borys Grinchenko Kyiv University*

# Analysis of Recent Attacks Based on Social Engineering Techniques

*Introduction.* The history of attacks based on SE practices is a wave: the victims changed, the different, new at their time, tricks were practiced and still are. The era of SE attacks in the field of IT began in 2014 when the first mass attacks were carried out on individuals, users of the banking payment system. People received calls from fake bank operators who informed about innovations regarding the protection of their data and steps that each and should every pass in order to become more secured. At their request, individuals in conversation gave critical data such as CVV2/CVC2 (3 digits on the back of a bank card) and 4 to 6 digits codes that the operator sent on their smartphones to confirm changes applying during the conversation, also in some cases even card pin-codes. The result of such manipulations as can understood had not given an additional level of protection to users, but rather deprived them of many decent sums of money (see fig. 1).

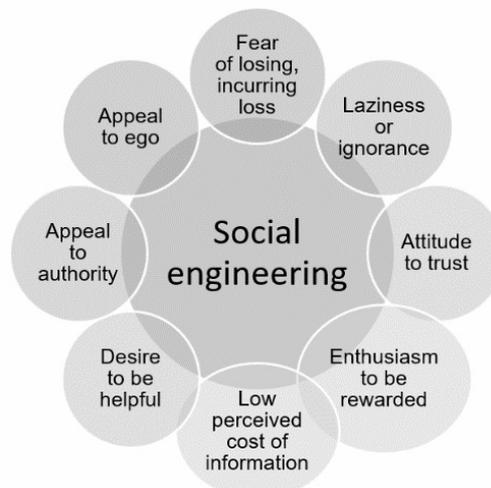

Figure 1 — Exploitation of human behavior

*1. Classification and description of known attack methods.* In 2015, European Central Bank deputy head of Security and Information Protection admitted that cybercriminals had switched to banks from their clients. The methods were not as advanced as can be seen now, mailing Trojan emails. A bank employee, opening such infected attachment "allows" an attacker to gain access to the account and send a payment order to or from the bank, so in that time the hacker group "Anunak" once attacked more than fifty banks and five payment systems in Ukraine and the countries of the former USSR and was able to steal about 1 billion dollars. Hacker billionaires were found proved guilty and convicted. Banks, in turn, increased protection [1].

Then the trend changed—cyber fraudsters became interested in small and medium-sized businesses, as there was more money on their accounts than on individuals, and protection might be weaker (if there is any) than in large organizations. An additional plus is that such companies often do not have a dedicated information security (IS) unit. As a result, it is enough for hackers to infect the accountant's computer in order to gain access to the accounts. This can be done in several ways that are analyzed further. Infection can occur through resources popular with financial workers. If criminals manage to compromise these sites, they turn into "hotbeds of infection", as they may contain on their pages a malicious exploit code (a subtype of malware). It uses an open browser vulnerability and establishes a "tunnel" with the user's computer. Through it, a program is loaded into the PC that determines what valuable information is stored on it. And then the "victim" is infected with a virus, specially adapted for antivirus on PC [2].

In 2016, targeted recruitment of insiders began to gain popularity. IS experts have declared the activity of intruders in this direction: more frequent attempts have been made to recruit bank employees, especially those who are part of the economic unit and are able to influence the adoption of certain decisions in the bank. A 2017 report by RedOwl and IntSights confirms the growing demand for insiders on the Dark Web. Employees are recruited purposefully, which greatly reduces the price of an attack: no need to guess how to penetrate the company's network and how to take out the data. For the "percentage from income" this information will provide an insider [3].

In 2017, silent ATM hacks began to gain popularity. When the device itself voluntarily gives money. To carry out such a crime without the help of insiders is extremely difficult. Criminals need information about the device ATM, the software built into it. And the test ATM modules (parts) itself, for training (fig. 2).

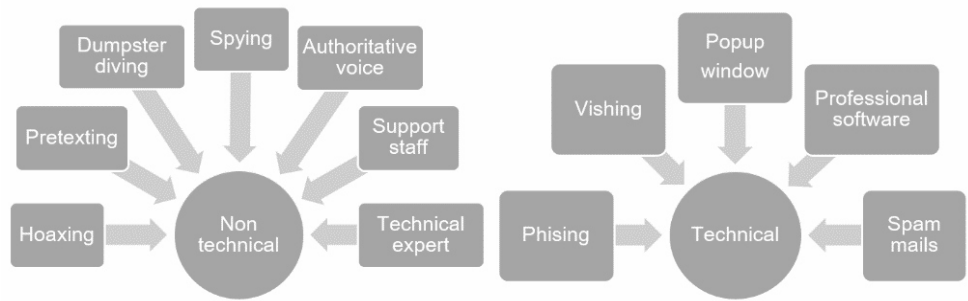

Figure 2 — Attack vectors

Previously, SE techniques united a common goal: the attacker caused obvious damage to the victim—obtaining information, financial damage, spoiled reputation, and demand for ransom. Therefore, it was exactly as long as the world was not overwhelmed with the fever of cryptocurrency mining. Mining promotes a simple idea for the masses: make money out of nothing. All you need to do is take a computer and use its power to "calculate" virtual currency. Currencies, by the way, are offered in abundance. In addition to replicated Bitcoin, today you can "invest with iron" in a dozen alternatives: Monero, litecoin, Zcash and others. But if everything was so simple, we would have been millionaires long ago. Using one computer, mining is economically unprofitable. For simplicity, the situation looks like this: earnings depend on how many hashes per second the processor or video card calculates (what exactly will be used depends on the specific cryptocurrency). For example, Monero is "calculated" by processors. With a performance of 863 kH/sec, you can earn $2,000 equivalent per day. That's just the performance of an Intel Core i5-7400 mid-range processor of about 0.165 kH/sec. This means that in a day at such capacities it will turn out to get as much as 38.3 cents [4].

The new goal of the census (social) engineers is to parasitize on the victim's technique. Of course, this created difficulties and led to an interesting effect—the goal of social engineers evolved. Now in 2018 the main task is not to cause obvious harm to the victim, but to quiet and inconspicuous parasitism on her technique. After all, the longer the virus miner will be on the car of an unsuspecting "donor", the more it "counts". Then revenues soar. Figures in confirmation: in the beginning of 2018 a group of hackers installed malware for the extraction of cryptocurrency on 9,000 computers via web-sites cookies and, according to analysts, such a network brings its owners up to $30,000 per month [5].

2019[th] may become the beginning of the era of "friendly" SE. Economically, mining "in the forehead" at its own expense is unprofitable (if not considering specialized devices and farms). Therefore, a new field of activity opens up for social engineers. In theory, mine cryptocurrency is possible on any device that has computing power and access to the Internet. Moreover, this is not only smartphones but also the whole range of IoT devices (or "smart devices"). In addition, for mining it is not necessary to install some kind of software, rather a special script. I think that this is the beginning of a new era—"undisguised" and "friendly" SE. And it is possible that soon, for example, banners will appear on torrent sites with a cat from Shrek and the words: "Please mine form two minutes. This will help us continue uploading pirated-movies for you." Honestly and without cheating the user.

*2. Pattern of changing SE threats.* There is a general principle "every action has a reaction" the more often attacks of the same type occur, the more identified companies and individuals become in the methods of struggle, prevention and further protect against them, the principle of SE implies that a person will always remain imperfect by creating, in certain circumstances, even a very savvy methodically person can suffer from the proper level of a trained attacker. The graph below shows that the society does not develop evenly known threats and people know about them, people have become more cautious, security policies are more strict and closed, but even now after almost 5 years from the first cases (fig. 3), even an obviously suspicious email can be opened and skipped by spam filter and antivirus and eventually opened by a computer user.

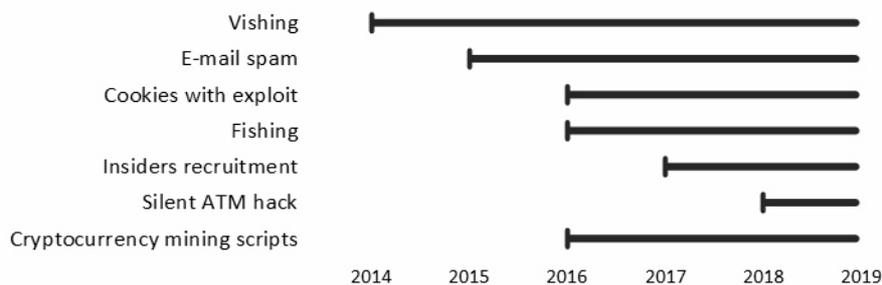

Figure 3 — Pattern of changing social engineering threats

*Concluding and further studies.* In this paper, there is only an attempt to outline the problem of social engineering. In the future, a full-scale research is planned on the reaction of people to phishing projects as part of the practice of ethical hacking.